\documentclass{aa}
\usepackage{graphics}
\usepackage{txfonts}

\newcommand{\mini}{\mbox{$M_{\rm i}$}}
\newcommand{\mto}{\mbox{$M_{\rm TO}$}}

\newcommand{\mbol}{\mbox{$m_{\rm bol}$}}
\newcommand{\Mbol}{\mbox{$M_{\rm bol}$}}

\newcommand{\feh}{\mbox{\rm [{\rm Fe}/{\rm H}]}}

\newcommand{\Msun}{\mbox{$M_{\odot}$}}
\newcommand{\Teff}{\mbox{$T_{\rm eff}$}}

\newcommand{\comment}[1]{}
\newcommand{\beq}{\begin{equation}}
\newcommand{\eeq}{\end{equation}}
\newcommand{\beqa}{\begin{eqnarray}}
\newcommand{\eeqa}{\end{eqnarray}}
        \def\smallskip{\vskip 2pt}

\begin{document}
\title{The TP-AGB phase. } 
\subtitle{Lifetimes from C and M star counts in Magellanic 
Cloud clusters}

\author{ L\'eo~Girardi$^1$ \and Paola Marigo$^2$}
\institute{
 Osservatorio Astronomico di Padova -- INAF, Vicolo dell'Osservatorio
 1, I-35122 Padova, Italy \and Dipartimento di Astronomia,
 Universit\`a di Padova, Vicolo dell'Osservatorio 2, I-35122 Padova,
 Italy }

\offprints{L. Girardi \\ e-mail: leo.girardi@oapd.inaf.it} 

\date{To appear in Astronomy \& Astrophysics}

\abstract{
Using available data for C and M giants with $\Mbol<-3.6$ in
Magellanic Cloud clusters, we derive limits to the lifetimes of the
corresponding evolutionary phases, as a function of stellar mass. The
C-star phase is found to have a duration between $2$ and $3$ Myr for
stars in the mass range from $\sim1.5$ to $2.8\,M_\odot$.  There is
also an indication that the peak of C-star lifetime shifts to lower
masses (from slightly above to slightly below 2~\Msun) as we move from
LMC to SMC metallicities. The M-giant lifetimes also peak at
$\sim2$~\Msun\ in the LMC, with a maximum value of about 4~Myr,
whereas in the SMC their lifetimes appear much shorter but, actually,
they are poorly constrained by the data. These numbers constitute
useful constraints to theoretical models of the TP-AGB phase. We show
that several models in the literature underestimate the duration of
the C-star phase at LMC metallicities.}

\authorrunning{L. Girardi \& P. Marigo}
\titlerunning{TP-AGB lifetimes}
\maketitle

\section{Introduction}
\label{intro}

Since the work by Frogel et al. (1990, hereafter FMB90), the thermally
pulsing asymptotic giant branch (TP-AGB) phase has been recognized to
be an important contributor to the integrated near-infrared luminosity
of young and intermediate-age stellar populations. Starting from the
observed numbers and luminosities of C and M-type giants in Magellanic
Cloud star clusters, FMB90 concluded that up to 40 percent of the
bolometric cluster luminosity comes from stars with $\Mbol<-3.6$,
which are above the RGB tip and hence belong to the TP-AGB. An
interesting plot by FMB90 showed this fraction as a function of SWB
cluster type (Searle et al. 1980), which gives approximate estimates
of cluster ages. Charlot \& Bruzual (1991) for the first time used
this information to calibrate the amount of TP-AGB stars added into
models of the spectrophotometric evolution of galaxies. This was an
important piece of information, previously missing, that has
contributed to the success of such models in interpreting the observed
spectra of galaxies.

Since then several different approaches have been adopted to include
the TP-AGB phase into evolutionary population synthesis models. They
often use slightly modified versions of the original FMB90 plots to
constrain the amount of TP-AGB stars (see Mouhcine \& Lan\c con 2002;
Maraston 2005, and references therein). Maraston (1998, 2005), for
instance, follows an empirical approach based on FMB90 data to add the
appropriate ``fuel consumption'' (equivalent to the contribution of
these stars to the integrated bolometric light) at LMC metallicities,
and then she uses results from uncalibrated TP-AGB evolutionary models
from Renzini \& Voli (1981) to account for the basic dependencies with
metallicity. Other authors directly include extended sets of TP-AGB
evolutionary models, for a wide enough range of stellar masses and
metallicities, into their models. This kind of approach goes from
adopting a very simplified description of the TP-AGB (e.g. Bressan et
al. 1994) to adding TP-AGB tracks {in which the efficiency of the
third dredge-up is} calibrated to reproduce the C-star luminosity
functions (CSLF) in the Magellanic Clouds (e.g. Marigo \& Girardi
2001). {Different schemes are adopted by Bruzual \& Charlot (2003)
and Mouhcine \& Lan\c con (2002), who adopt TP-AGB tracks that
distinguish between the C- and M-type phases, but which are not
directly calibrated on the CSLF: Bruzual \& Charlot (2003) make a
composition of TP-AGB tracks from different sources (namely
Vassiliadis \& Wood 1993, Groenewegen \& de Jong 1993, and Groenewegen
et al. 1995), but do not check the consistency of these composite
tracks with the observed properties of resolved AGB
populations. Mouhcine \& Lan\c con (2002) instead compute their own
TP-AGB tracks adopting modified prescriptions but the same values of
dredge-up parameters as Groenewegen \& de Jong (1993); they make
several different comparisons with data, avoiding however the explicit
check with the CSLFs that was key to Groenewegen \& de Jong (1993).}

Owing to the many interrelated aspects of TP-AGB evolution and their
complex dependence on metallicity and stellar mass, our understanding
is that the best approach to be used in population synthesis should be
the inclusion of TP-AGB tracks computed in a self-consistent way and
{\em directly} calibrated using a set of observables from the LMC and
SMC. Once calibrated, these self-consistent sets of TP-AGB tracks
present the potential advantage of better describing the dependence of
all TP-AGB properties on the stellar metallicity, and hence they
should work better for metallicities significantly different from the
LMC and SMC ones. In this context, the main goal of the present paper
is to translate FMB90 data into useful quantities -- the TP-AGB
lifetimes as a function of stellar mass -- for the direct calibration
of TP-AGB tracks of LMC ad SMC metallicities.

Originally, this work has been motivated by the finding by Marigo
(2002) that all TP-AGB models computed that far have used a very
improper prescription for their low-temperature opacities. She
replaced the opacities for solar-scaled chemical mixtures -- still now
the standard choice in the literature -- by opacities properly
evaluated for the chemical mixtures of evolving TP-AGB envelopes.  One
main evolutionary effect of variable molecular opacities is the
remarkable reduction of effective temperatures for C stars, which then
causes an earlier onset of the superwind regime and a reduction of
their lifetimes, when compared to models computed at fixed
solar-scaled opacities. This reduction should not affect the previous
M-type phase, for which no dramatic change in molecular opacities is
expected. These findings also cause serious doubts about the
previously-derived behaviours of TP-AGB evolutionary properties --
including lifetimes, termination luminosities, effective temperatures,
etc -- with metallicity. It is then evident that new grids of TP-AGB
tracks are due, and that the C-star lifetimes, once derived from
empirical data, could provide important constraints to them.

\section{Cluster data}

Magellanic Clouds young clusters are clearly the best objects to check
TP-AGB lifetimes by means of C- and M-star counts: They are populous
enough to contain appreciable numbers of cool giants, have already
been searched for them (FMB90 and references therein), and at the same
time they have reasonably well-known distances, ages and
metallicities. In order to relate the observed numbers of C and M
stars to their lifetimes, we also need a measure of each cluster's
size. The total masses are very uncertain even for the best studied
clusters, and are conditioned by the large mass fraction locked up in
low-mass dwarfs. In practice, a better measure of cluster's size comes
from its integrated luminosities. In this paper, we deal with the
integrated $V$-band luminosity, $L_V$, for a series of reasons: First
the integrated $V$ magnitudes are known for all clusters in FMB90's
catalog; second $L_V$ smoothly declines with cluster age and it is
expected to be little sensitive to errors in the cluster metallicity
(cf. Girardi 2000); and third $L_V$ samples stars in well-populated
evolutionary stages (namely close to the main-sequence turn-off and
core-He burning, see Charlot \& Bruzual 1991; Girardi \& Bica 1993)
and hence it is little affected by stochastic cluster-to-cluster
variations in their number of stars. In comparison, the integrated
luminosities in red and near-IR passbands, like $I$ and $K$, although
available from wide-area surveys such as 2MASS and DENIS, are too
sensitive to the stars in the upper part of the RGB and AGB. They
present significant non-monotonic behaviours with both age and
metallicity, and stochastic cluster-to-cluster variations (Girardi
2002), which we prefer to avoid.

Therefore, a good starting point to derive lifetimes is to use the
observed total number of C- and M-stars in a cluster, $N_{\rm C}$ and
$N_{\rm M}$, divided by its integrated $V$-band luminosity,
$L_V$. These observed quantities are directly proportional to the M
and C-type lifetimes. For the M stars, we limit the comparison to the
AGB stars above the RGB-tip, i.e. those with $\Mbol<-3.6$. Adopting
LMC and SMC distance moduli of 18.5 and 18.9 mag, this limit
correspond to entries of $\mbol<14.9$ and $\mbol<15.3$, for LMC and SMC
stars respectively, listed in the table~1 of FMB90.

\begin{table*}
\caption{C and M data for Magellanic Cloud clusters} 
\label{tab_cluster}
\begin{tabular}{lrrlrrrr}
\hline\hline
Id. & $\log(t/{\rm yr})$ & $S$ & { \feh} & $N_{\rm C}$ & $N_{\rm M}$ & $V$ & 
$\sigma_{\rm C}$\\
\hline
LMC: \\
NGC~1854 & 7.66	& 22 & -- & 0 & 2 & 10.39 & 300--500\\
NGC~1850 & 7.78	& 23 & -- & 1 & 4 &  9.57 & 300--500\\
NGC~2214 & 7.78	& 23 & -- & 0 & 2 & 10.93 & 25--75\\
NGC~2136 & 7.78 & 25 & -0.55$^{\rm e}$ & 0 & 0 & 10.54 & 150--300\\
NGC~2058 & 8.03 & 25 & -- & 0 & 5 & 11.85 & 600\\ 
NGC~1866 & 8.08 & 28 & -0.50$^{\rm f}$,-0.55$^{\rm d}$ & 0 & 3 &  9.73 & 25--75\\
NGC~2107 &   -- & 32 & -- & 0 & 1 & 11.51 & 300--500\\
NGC~1987 &   -- & 35 & -0.50$^{\rm f}$,-0.50$^{\rm d}$ & 1 & 3 & 12.08 & 300\\
NGC~2209 & 9.03 & 35 & -- & 2 & 0 & 13.15 & 25\\
NGC~2108 &   -- & 36 & -- & 1 & 1 & 12.32 & 150--300\\
NGC~1783 &   -- & 37 & -0.75$^{\rm c}$ & 4 &10 & 10.93 & 25--75\\ 
NGC~2213 & 8.99 & 38 & -0.01$^{\rm a}$ & 3 & 1 & 12.38 & 25--75\\
NGC~2231 & 9.26 & 38 & -0.52$^{\rm h}$,-0.67$^{\rm a}$ & 1 & 1 & 13.20 & 25--75\\
NGC~2154 &   -- & 38 & -0.56$^{\rm a}$ & 2 & 2 & 11.79 & 25--75 \\
NGC~1806 &   -- & 38 & -0.71$^{\rm e}$,-0.23$^{\rm a}$ & 2 & 6 & 11.10 & 150--300\\
NGC~1651 & 9.24 & 38 & -0.53$^{\rm h}$,-0.53$^{\rm e}$,-0.37:$^{\rm a}$ & 1 & 3 & 12.28 & 25 \\
NGC~1846 &   -- & 39 & -0.49$^{\rm h}$,-0.70$^{\rm a}$ & 9 & 9 & 11.31 & 150  \\
NGC~1751 &   -- & 40 & -0.44$^{\rm h}$,-0.18:$^{\rm a}$ & 2 & 4 & 11.73 & 150 \\
NGC~1652 &   -- & 41 & -0.46$^{\rm h}$,-0.45$^{\rm a}$ & 0 & 0 & 13.13 & 25 \\
NGC~1978 & 9.40 & 41 & -0.38$^{\rm g}$,-0.96$^{\rm f}$,-0.60$^{\rm c}$,-0.41$^{\rm a}$ & 6 & 3 & 10.70 & 25--75 \\ 
NGC~2173 & 9.18 & 41 & -0.42$^{\rm h}$,-0.50$^{\rm c}$,-0.24$^{\rm a}$ & 1 & 3 & 11.88 & 25--75  \\
NGC~2121 & 9.03 & 46 & -0.50$^{\rm h}$,-0.10$^{\rm c}$,-0.61:$^{\rm a}$ & 0 & 2 & 12.37 & 150 \\ 
NGC~1841 & 9.90 & 54 & -2.02$^{\rm h}$ & 0 & 0 & 11.43 & $<$25 \\
\hline
SMC: \\
NGC~416 & 8.78  & 35 & -0.80$^{\rm c}$ & 1 & 0 & 11.42 & 150--300\\ 
NGC~419 & 9.08  & 39 & -0.60$^{\rm c}$ &10 & 0 & 10.61 & 150--300\\ 
NGC~411 & 9.26  & 41 & -0.70$^{\rm c}$ & 2 & 0 & 12.21 & 75--150\\
NGC~152 & 9.28  & 42 & -- & 2 & 1 & 12.92 & 150\\
Kron~3  & 9.67  & 47 & -1.00$^{\rm c}$,-0.98$^{\rm b}$ & 3 & 0 & 12.05 & 10--75 \\
NGC~339 & 9.70  & 47 & -0.70$^{\rm c}$,-1.19$^{\rm b}$ & 1 & 0 & 12.84 & 75\\
NGC~361 & 9.83  & 49 & -- & 0 & 1 & 12.78 & 75\\
NGC~121 & 10.03 & 52 & -1.19$^{\rm b}$ & 1 & 1 & 11.24 & 10\\
\hline
\end{tabular}
\\ {
References for \feh: $^{\rm a}$ Olszewski et al (1991); $^{\rm b}$ Da
Costa \& Hatzidimitriou (1998) in the Carretta \& Gratton (1997)
scale; $^{\rm c}$ de Freitas Pacheco et al. (1998); $^{\rm d}$ Oliva
\& Origlia (1998); $^{\rm e}$ mean value from Dirsh et al. (2000);
$^{\rm f}$ Hill et al. (2000); $^{\rm g}$ Ferraro et al. (2006);
$^{\rm h}$ Grocholski et al (2006).}
\end{table*}

\begin{table*}
\caption{Final binned data for C and M stars in Magellanic Cloud
clusters} 
\label{tab_binned}
\begin{tabular}{lrrrrrrrrrr}
\hline\hline
$S$ interval & $\log t$  & \feh & $M_{\rm TO}$ & 
	$N_{\rm C}$ & $N_{\rm M}$ & $L_V$  &
	$N_{\rm C}/L_V$ & $N_{\rm M}/L_V$ & $\tau_{\rm C}$ & $\tau_{\rm M}$ \\
             & ($t$ in yr) & & (\Msun) & 
	 & & ($10^6\,L_{V\odot}$)  &
	($10^{-5}$) & ($10^{-5}$) & (Myr) & (Myr) \\
\hline
 LMC S22-24 &  7.91 & -0.20 &  5.90 &  1 &  8 & 6.23 & 
	$0.16_{-0.13}^{+0.37}$ & $1.28_{-0.44}^{+0.63}$ &
 	$0.032_{-0.026}^{+0.075}$ & $0.26_{-0.09}^{+0.13}$ \\ 
 LMC S25-27 &  8.13 & -0.20 &  4.75 &  0 &  5 & 1.90 & 
	$<0.60$ & $2.63_{-1.13}^{+1.77}$ &
	$<0.50$ & $2.19_{-0.94}^{+1.47}$ \\ 
 LMC S28-30 &  8.35 & -0.21 &  3.85 &  0 &  3 & 3.07 & 
	$<0.37$ & $0.98_{-0.53}^{+0.95}$ &
	$<0.07$ & $0.19_{-0.10}^{+0.18}$ \\ 
 LMC S31-33 &  8.57 & -0.24 &  3.17 &  0 &  1 & 0.60 & 
	$<1.90$ & $1.67_{-1.38}^{+3.81}$ &
	$<0.90$ & $0.79_{-0.66}^{+1.81}$ \\ 
 LMC S34-36 &  8.79 & -0.27 &  2.66 &  4 &  4 & 0.76 & 
	$5.26_{-2.51}^{+4.14}$ & $5.26_{-2.51}^{+4.14}$ &
	$2.80_{-1.34}^{+2.21}$ & $2.80_{-1.34}^{+2.21}$ \\ 
 LMC S37-39 &  9.01 & -0.35 &  2.17 & 22 & 32 & 3.76 & 
	$5.85_{-1.23}^{+1.52}$ & $8.51_{-1.49}^{+1.78}$ &
	$2.59_{-0.55}^{+0.67}$ & $3.77_{-0.66}^{+0.79}$ \\ 
 LMC S40-42 &  9.23 & -0.54 &  1.66 &  9 & 10 & 2.31 & 
	$3.90_{-1.27}^{+1.77}$ & $4.33_{-1.34}^{+1.84}$ &
	$1.57_{-0.51}^{+0.71}$ & $1.74_{-0.54}^{+0.74}$ \\ 
 LMC S46-48 &  9.67 & -0.60 &  1.18 &  0 &  2 & 0.27 & 
	$<4.22$ & $7.41_{-4.77}^{+9.72}$ &
	$<1.49$ & $2.62_{-1.69}^{+3.44}$ \\ 
 LMC S52-54 & 10.11 & -1.62 &  0.82 &  0 &  0 & 0.64 & 
	$<1.78$ & $<1.78$ &
    	$<0.66$ & $<0.66$ \\ 
 SMC S34-36 &  8.79 & -0.56 &  2.52 &  1 &  0 & 0.94 & 
	$1.06_{-0.88}^{+2.43}$ & $<1.21$ &
	$0.54_{-0.45}^{+1.23}$ & $<0.61$\\ 
 SMC S37-39 &  9.01 & -0.57 &  2.09 & 10 &  0 & 1.97 & 
	$5.08_{-1.57}^{+2.15}$ & $<0.58$ &
	$2.37_{-0.73}^{+1.00}$ & $<0.27$ \\ 
 SMC S40-42 &  9.23 & -0.72 &  1.62 &  4 &  1 & 0.68 & 
	$5.88_{-2.80}^{+4.63}$ & $1.47_{1.22}^{3.37}$ &
	$2.47_{-1.18}^{+1.95}$ & $0.62_{0.51}^{1.42}$ \\ 
 SMC S46-48 &  9.67 & -1.18 &  1.12 &  4 &  0 & 0.78 & 
	$5.13_{-2.45}^{+4.03}$ & $<1.46$ &
	$1.97_{-0.94}^{+1.55}$ & $<0.56$ \\ 
 SMC S49-51 &  9.89 & -1.23 &  0.96 &  0 &  1 & 0.27 & 
	$<4.22$ & $3.70_{-3.06}^{+8.48}$ &
	$<2.33$ & $2.05_{-1.69}^{+4.69}$ \\ 
 SMC S52-54 & 10.11 & -1.35 &  0.84 &  1 &  1 & 1.10 & 
	$0.91_{-0.75}^{+2.08}$ & $0.91_{-0.75}^{+2.08}$ &
	$0.32_{-0.26}^{+0.72}$ & $0.32_{-0.26}^{+0.72}$ \\ 
\hline
\end{tabular}
\end{table*}

Table \ref{tab_cluster} summarises the cluster data available to our
purposes. {We have considered all clusters in FMB90, excluding the
very young ones (i.e. those with $t\la10^8$~yr) and a few SMC clusters
for which we did not find age determinations based on the main
sequence turn-off photometry.} For each cluster, the entries in the
table correspond to:
\begin{itemize} 
\item The age $t$ as derived from main-sequence turn-off 
photometry, if available. It is taken from Girardi et al. (1995) for
LMC clusters, and Da Costa \& Hatzidimitriou (1998) and Mighell et
al. (1998) for SMC clusters.
\item The age-parameter $S$, as determined and calibrated by Girardi 
et al. (1995) for LMC clusters only. For the most populous young and
intermediate-age LMC objects (including the entries in
Table~\ref{tab_cluster}), the relation $\log(t/{\rm yr}) =
6.227+0.0733\,S$ gives the turn-off age $t$ with an error of about
$0.15$~dex in $\log t$. For SMC clusters, we derive the $S$ parameters
directly from the turn-off age using the same relation as for LMC
clusters.
\item {Recent determination of \feh\, when available from either 
spectroscopic data, or from methods directly calibrated with
spectroscopy; this \feh\ list is certainly very heterogeneous and
likely incomplete.}
\item The C-star counts, and M-star counts above $\Mbol<-3.6$, 
$N_{\rm C}$ and $N_{\rm M}$. They include the spectroscopically
confirmed C- and M-stars considered to be cluster members -- i.e.
located within a circle of diameter ``close to or somewhat larger than
$1$~arcmin'' around each cluster -- in FMB90's table 1, and the few
objects with dubious membership (``Y?'' in his column 4) or spectral
classification (``C?'' or ``M?''  in his column 3). Clusters without
such stars are also included in the table.
\item The cluster integrated $V$-band magnitude from Bica et al. (1996) 
for the LMC, and van den Bergh (1981) for the SMC. This can be easily
converted into the integrated luminosity in solar units, after
assuming the Sun has a $V$-band magnitude of $M_{V\,\odot}=4.847$, and
apparent distance moduli of 18.6 and 19.0 mag, for the LMC and SMC
respectively.
\item The surface density of field C-stars, $\sigma_{\rm C}$ in 
units of stars per deg$^2$, at the cluster position, as derived by
Blanco \& McCarthy (1983).  We give just the approximate position of
each cluster in these isopleth maps (their figures 2 and 3).
\end{itemize}

\begin{figure*}  
\begin{minipage}{0.47\textwidth}
	\resizebox{\hsize}{!}{\includegraphics{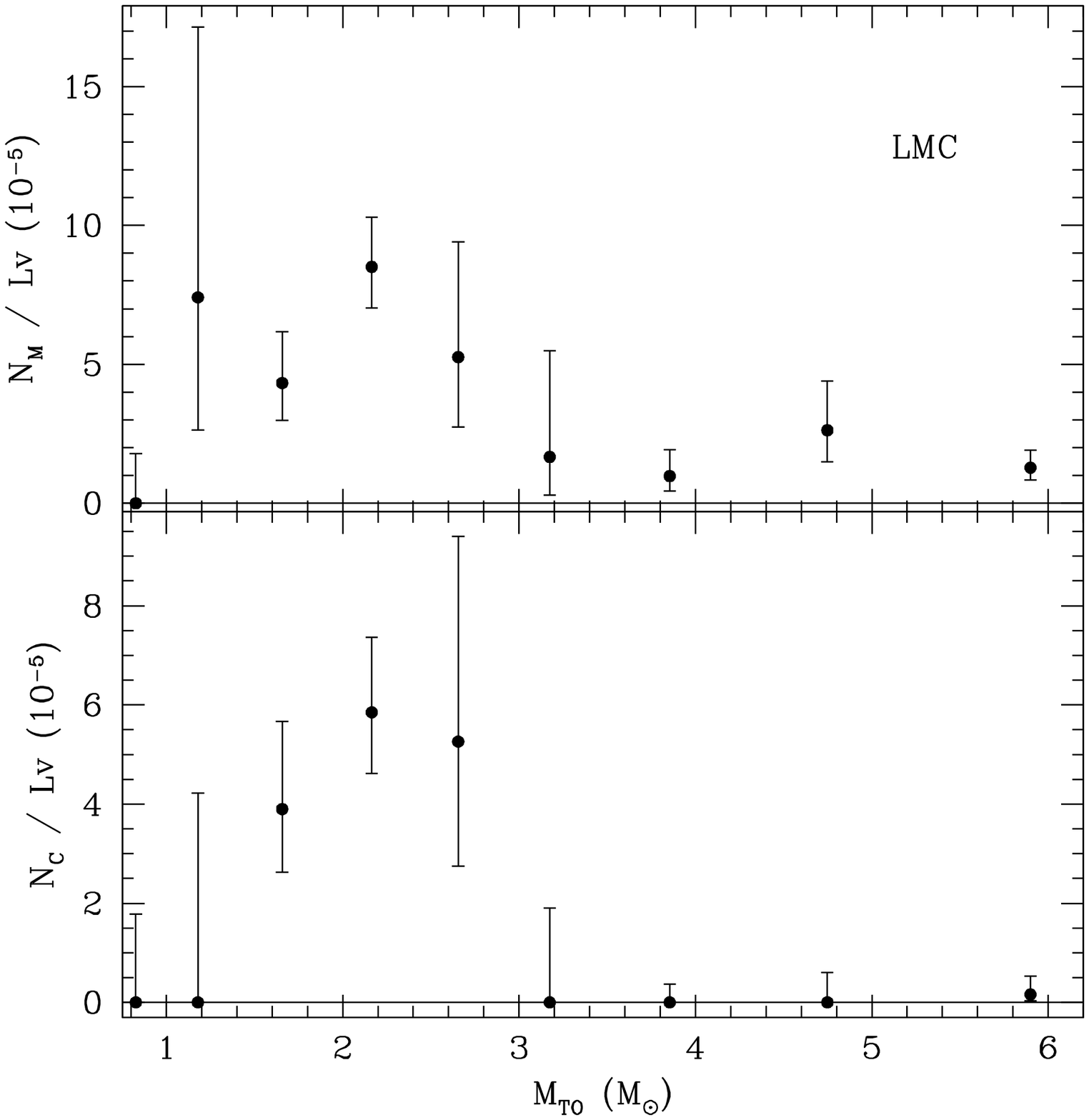}}
\end{minipage} 
\hfill
\begin{minipage}{0.47\textwidth}
	\resizebox{\hsize}{!}{\includegraphics{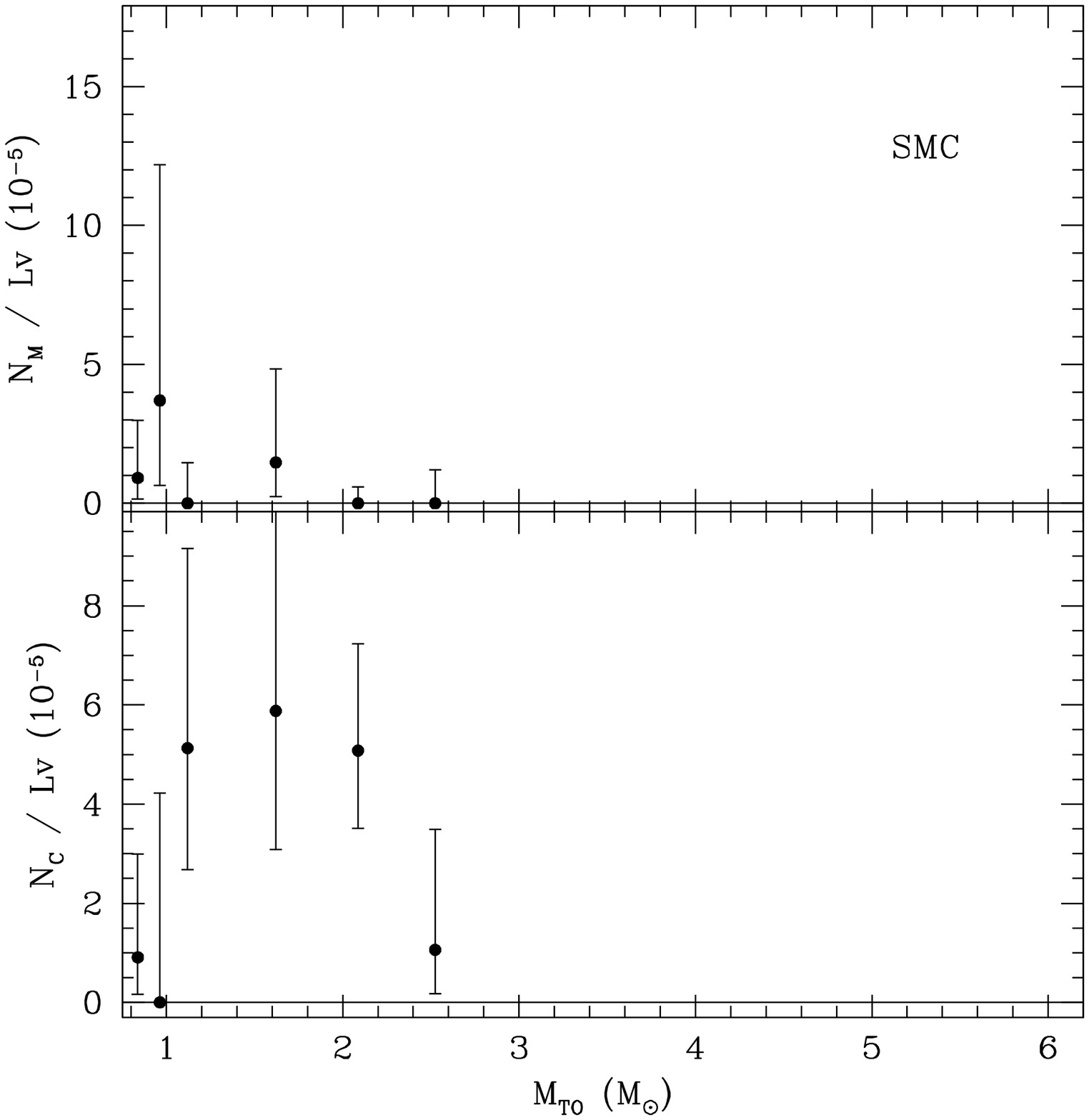}}
\end{minipage} 
\caption{
The ratio between the number of C and M stars in clusters, and their
integrated $V$-band luminosity, $N_{\rm C}/L_V$ and $N_{\rm M}/L_V$
(bottom and top panels, respectively), as a function of turn-off
mass. The data are shown separately for the LMC (left panel) and SMC
(right). }
\label{fig_nl}
\end{figure*}

For the SMC clusters NGC~419 and NGC~152, FMB90 catalogue contains
many member stars without spectral classification but bright enough
($\mbol<15.3$) to be in the TP-AGB phase. For the sake of homogeneity,
we do not include these objects. We suspect they correspond to O-rich
TP-AGB stars in these clusters, which would otherwise be practically
missing.  This fact warn us that the data for the O-rich TP-AGB stars
above $\Mbol=-3.6$ may be severely incomplete for the SMC. This does
not happen for the LMC, where, owing to smaller distance (and maybe to
the slightly larger mean metallicities and smaller \Teff), stars
slightly above $\mbol<14.9$ are, as a rule, clearly classified by
FMB90 as being of early M subtypes (M0--M2).

The reader will also notice the high degree of standardization we
applied to these data, like a single value of distance+reddening for
all clusters in each galaxy, and the approximative ages. They are
necessary because, for the moment, there seems to be no satisfactorily
homogeneous and updated compilation of such data. Apart from these
problems, it is immediately evident from Table~\ref{tab_cluster} that,
in the end, the AGB data is very scarce when we consider individual
clusters (see also Marigo et al.\ 1996). Just a handful of LMC
clusters host more than two confirmed C stars, whereas for the SMC
there are just two of such clusters. Taken cluster-per-cluster, the
statistics provided by these data would be very poor. Therefore, we
have added together the cluster data in bins of $\Delta S=3$, which
corresponds to age bins of $\Delta\log t=0.22$ and turn-off mass bins
of $\Delta\log M_{\rm TO}\simeq-0.10$.

The binned data are presented in Table~\ref{tab_binned}, together with
the expected mean metallicity \feh\ for each bin as given by Pagel \&
Tautvaisiene's (1998) age--metallicity relations (AMR) for ``bursting
models'' of the LMC and SMC. These AMRs fit well the \feh\ data of
individual LMC and SMC clusters, within the observed scatter of about
0.2~dex ($1\sigma$) for a given age. Finally, Table~\ref{tab_binned}
presents the estimated turn-off mass, $\mto$, for the middle of each
bin, as derived from its mean $\log t$ and \feh, and Girardi et
al. (2000) evolutionary tracks.  The mass of the more evolved AGB
stars should be slightly higher than \mto\ but still very close to the
tabulated values, with maximum differences amounting to just
$\sim0.1$~\Msun.

How large is the contamination by C and M stars belonging to the LMC
and SMC fields rather than to the clusters\,? We look for the answer
in Blanco \& McCarthy's (1983) study of the surface distribution of C-
and M-type giants across the Magellanic Clouds. Their figures 2 and 3
show isopleths of field C-star densities, $\sigma_{\rm C}$, over both
galaxies; the last column of Table~\ref{tab_cluster} reports the
location of FMB90 clusters in these isopleth maps. The highest values
of $\sigma_{\rm C}$, of the order of 600~deg$^{-2}$, are found just in
the central SMC and LMC bar regions. NGC~2058 is the only cluster in
our sample located in such a high-density field; for it, the
$\sigma_{\rm C}=600$~deg$^{-2}$ value would translate into an expected
number of $\sim0.13$ field C stars inside the
$\sim2\times10^{-4}$~deg$^{-2}$ area surveyed by FMB90. This is
already a very low expectancy value.  Most of the clusters in
Table~\ref{tab_cluster}, however, are in outer LMC and SMC regions
with $\sigma_{\rm C}$ values well below 300~deg$^{-2}$. Adding all the
individual $\sigma_{\rm C}$ values multiplied by each cluster area,
the {\em total} number of contaminating C stars is estimated to be
comprised between 0.75 and 1.4. Compared to our total sample of 56
C-stars distributed in 31 clusters, this contamination is small enough
to be neglected. Even more importantly, the only individual clusters
with a significant probability of being contaminated by field C stars,
i.e. those with $\sigma_{\rm C}>300$~deg$^{-2}$, are young LMC
clusters with $S<32$, corresponding to turn-off masses higher than
3~\Msun. This is exactly the age interval with less observed C-stars
for which, as we will see later, just upper limits to the C-star
lifetimes can be derived. Therefore, contamination by field C stars
seems not to be a problem\footnote{Cioni \& Habing (2003), using
DENIS data, find a total number of 7750 C-type stars in the LMC, less
than the $\sim11000$ expected from Blanco \& McCarthy's (1983)
maps. This would be reassuring because the field C-star contamination
could be even smaller than here estimated. However, we remark that the
Cioni \& Habing (2003) classification is based on a photometric
criterion that likely misclassifies a non-negligible fraction
($\sim10~\%$) of C-type stars as M-type ones, therefore there may be
no real discrepancy in their estimates when compared to Blanco
\& McCarthy's (1983) one. In this paper, we opt to use the Blanco
\& McCarthy (1983) C-star numbers because they are based on a
spectroscopic classification which is equivalent to the one used by
FMB90, and which is considered to be complete for the C stars above
the RGB-tip.}.

Regarding the M-giants, Blanco \& McCarthy (1983) data cannot be used
to estimate the field contamination since it is complete only for
spectral types later than M5 (i.e. M5+). Cioni \& Habing (2003, their
table 1), from completely independent data, find the C/M0+ ratio for
stars above the RGB tip to be 0.30 and 0.27, for the entire LMC and
SMC, respectively.  From these numbers, we can roughly estimate that
the contamination by field M-giants is just a few times larger {
($\sim3.3$ and $\sim3.7$ times, for the LMC and SMC) than the one from
C-giants. Assuming 3.5 as the overdensity factor for both LMC and SMC,
we can expect from 2.5 to 5 field M-stars contaminating our sample,
which contains 68 M stars in total (65 in the LMC and 3 in the
SMC). We conclude that the field M-giant contamination can be safely
neglected as well, at least for the LMC. For the SMC, the total number
of bona-fide M giants in clusters (just 3) is so low that field
contamination may indeed be an issue; however, as we conclude later in
Sect.~\ref{sec_conclu}, this number is also low enough to make the SMC
M-star data almost useless as a constraint to AGB models.}

The FMB90 data refers only to optically visible stars.  What about the
presence of dust-enshrouded, optically obscured TP-AGB stars? van Loon
et al. (2005) present a comprehensive survey and investigation of such
stars in Magellanic Cloud Clusters, thus representing the mid-infrared
counterpart of FMB90 survey. Unfortunately, the two cluster samples
are quite different. van Loon et al.'s (2005) sample of bright IR
cluster objects (their tables 2 to 7) would add stars to just three of
our 31 clusters, namely: 1 AGB C-star in NGC~1783, 1 AGB C-star and 1
(post-)AGB C-star to NGC~1978, and 2 AGB C-stars for NGC~419. It is
clear that dust-obscured objects in clusters constitute a modest
fraction of the total numbers of TP-AGB stars. We will come back to
this point later.

The final $N_{\rm C}/L_V$ and $N_{\rm M}/L_V$ data are presented in
Table~\ref{tab_binned} and Figure~\ref{fig_nl} together with upper and
lower limits given by the 68\% confidence level interval of a Poisson
distribution (i.e. $1\sigma$ for the most populated bins).

\section{Lifetimes as a function of mass}

\begin{figure*}  
\begin{minipage}{0.47\textwidth}
	\resizebox{\hsize}{!}{\includegraphics{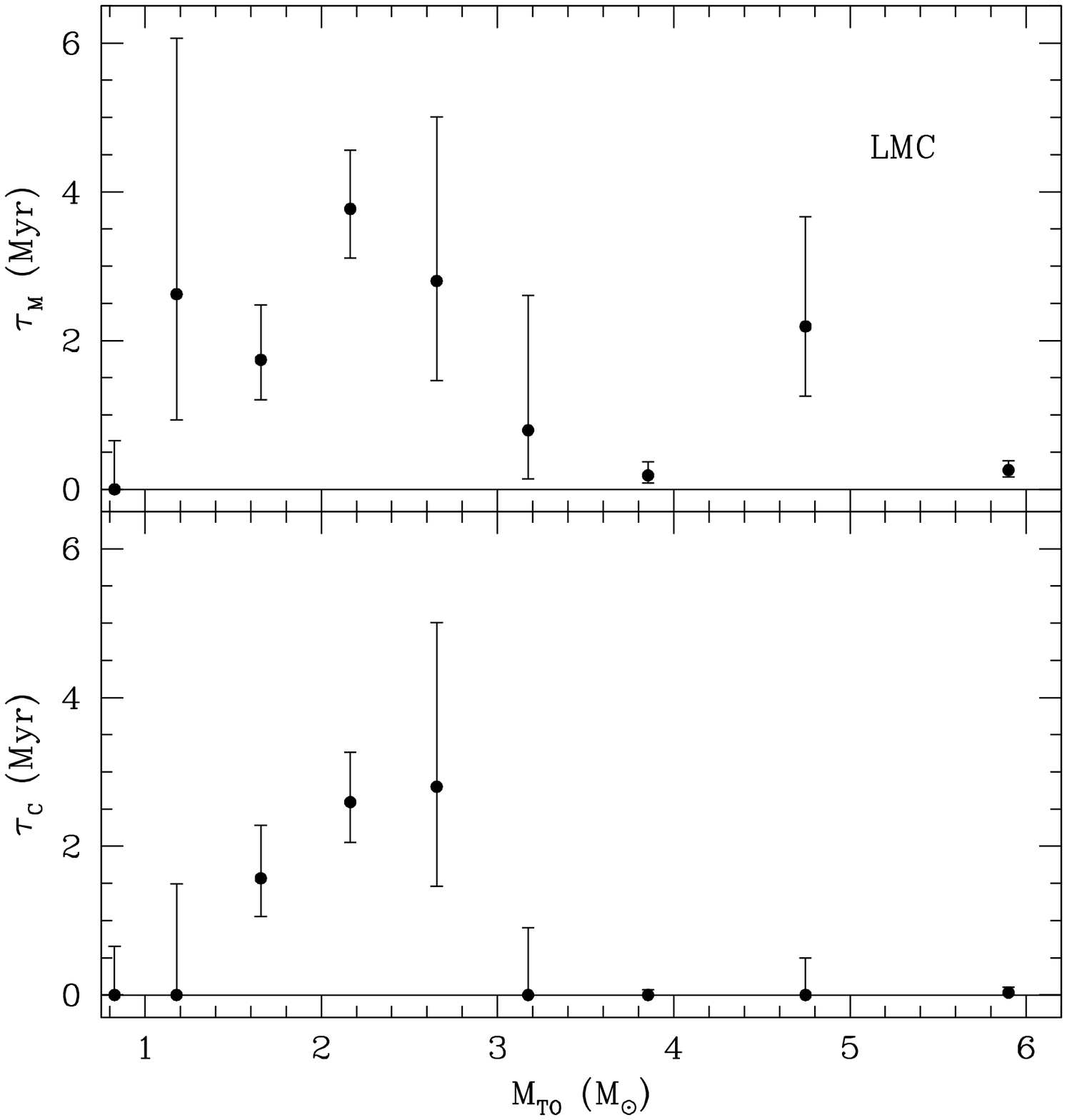}}
\end{minipage} 
\hfill
\begin{minipage}{0.47\textwidth}
	\resizebox{\hsize}{!}{\includegraphics{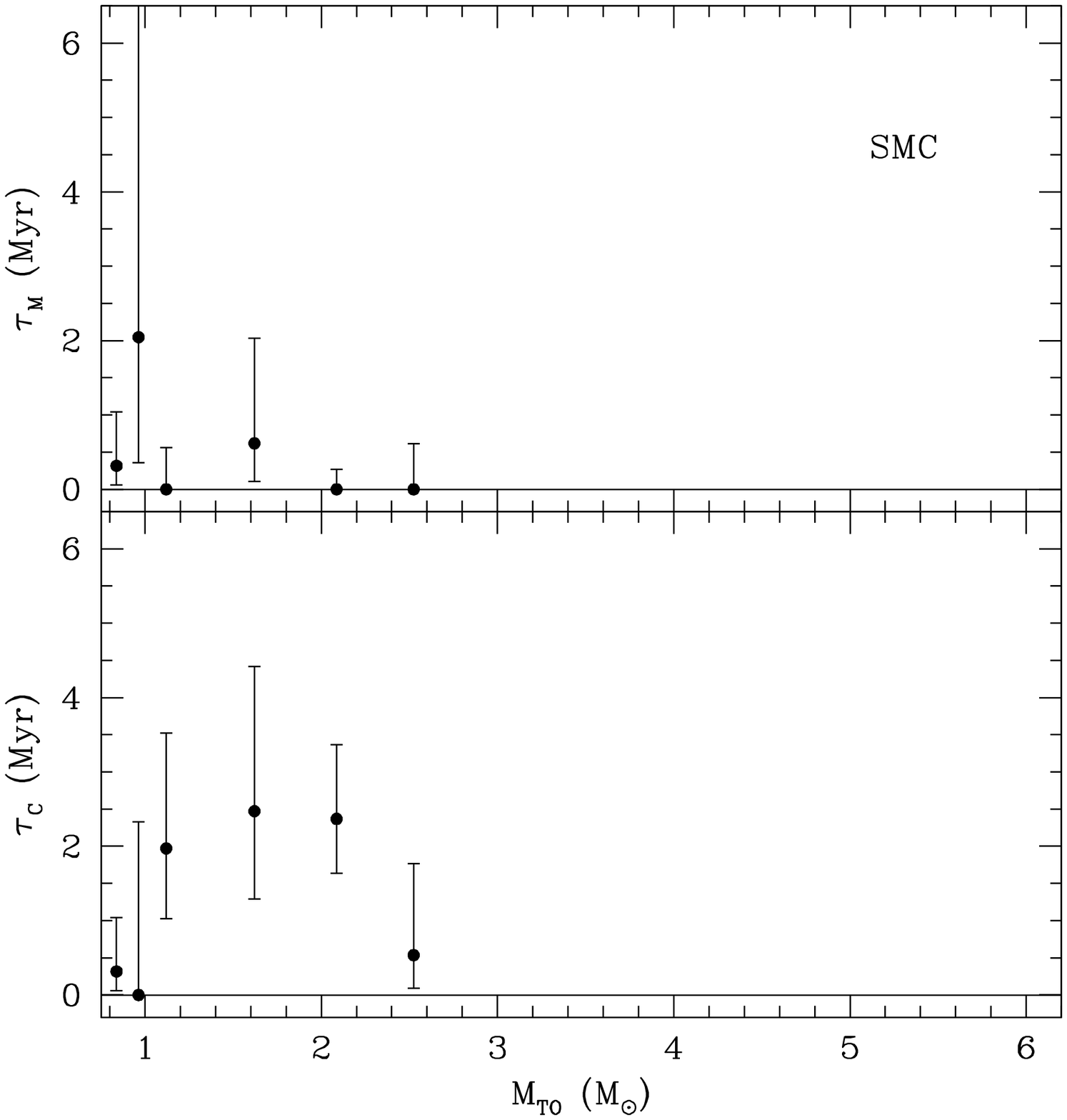}}
\end{minipage} 
\caption{
The lifetimes of the C- and M-type phases (limited to $\Mbol<-3.6$ in
the case of M stars) as inferred from the cluster data, as a function
of turn-off mass. The data are shown separately for the LMC (left
panel) and SMC (right).}
\label{fig_lifetimes}
\end{figure*}

The $N_{\rm C}/L_V$ and $N_{\rm M}/L_V$ quantities in
Table~\ref{tab_binned} could already be directly used to constrain
theoretical models, since they are proportional to the typical C- and
M-type lifetimes.  Large error bars are implied by the low number of
available objects for some of the age bins. Anyway, the LMC data pose
well defined constraints to the C-type lifetime for masses between 1.5
and 2.8~$M_\odot$, whereas at higher/lower masses just upper limits
are derived. For the SMC, error bars are even larger, but three age
bins have enough statistics to provide useful constraints to $N_{\rm
C}/L_V$ between 1.2 and 2.4~$M_\odot$. The $N_{\rm M}/L_V$ data for
the SMC is in general of low quality, and it is likely severely
affected by incompleteness; moreover, at SMC metallicities a
significant fraction of the O-rich AGB stars above the RGB-tip may be
of spectral type earlier than M, and then they may be absent from
FMB90 catalog. As a consequence, one should better not using the
$N_{\rm M}/L_V$ ratios derived in the SMC to constrain AGB models.

The $N/L_V$ values can now be converted directly into stellar
lifetimes as follows. For a given evolutionary stage $j$, $N_j/L_V$ is
related to the the lifetime $\tau_j$ through an age-dependent
proportionality constant, that could be computed by using basic
population synthesis theory. In this paper, we compute the
proportionality constant in a purely numerical way: we take a set of
AGB models whose AGB lifetimes $\tau_{\rm AGB}(\mini,Z)$ are exactly
known for all masses and metallicities. Then, we construct isochrones
and, by simply integrating the stellar number density and $V$-band
luminosity, {weighted by the initial mass function (IMF)}, along
them\footnote{This integration is performed assuming Kroupa's (2001)
IMF corrected for binaries, but the results are quite insensitive to
the IMF.}, we derive the theoretical isochrone $N_{\rm AGB}/L_V$
ratios. The empirical lifetimes of the phase $j$ under consideration
are then given by the ratio between the observed $N_j/L_V$
(Table~\ref{tab_binned}) and the corresponding simulated $N_{\rm
AGB}/L_V$, times the $\tau_{\rm AGB}(\mini,Z)$ lifetime that
corresponds to the AGB stars in that isochrone.  By interpolating
between models of several masses and metallicities, we properly take
into consideration the variation with metallicity of the main sequence
lifetimes -- and hence of the evolutionary rate at which stars leave
the main sequence -- and of the integrated $V$-band luminosity. Both
effects play a non negligible role in determining the proportionality
constant between $N_j/L_V$ and $\tau_j$.

The final results for the lifetimes of C and M-type giants as a
function of mass are presented in the last columns of
Table~\ref{tab_binned} and in Fig.~\ref{fig_lifetimes}.

{ In the present work, we have used the same isochrones as in Cioni
et al. (2006ab)}\footnote{These isochrones are available in {\tt
http://pleiadi.oapd.inaf.it}}; { they are based on Girardi et
al. (2000) tracks for the pre-TP-AGB phases, and completed with TP-AGB
tracks computed on purpose using Marigo's (2002) code. As a matter of
fact, the $L_V$ values we have used depends very little on the
particular set of TP-AGB tracks used in the isochrones. We have
checked that excluding the TP-AGB from Cioni et al. (2006ab)
isochrones causes the integrated $V$-band magnitudes to increase by
less than 0.1~mag at all ages and metallicities relevant to this work;
0.04~mag is the typical value for this difference. This means that the
possible ``systematic'' errors in the lifetimes, caused by possible
errors in our TP-AGB tracks, are of the order of just $\sim4$~\%. This
is much smaller than the errors caused by the poor statistics in the
data (see the 68~\% confidence-level error bars in Fig.~\ref{fig_nl}),
so that we consider this problem as being of minor importance.}

{More relevant may be the errors caused by uncertainties in the
AMRs: at a given age and for both the LMC and SMC, the mean cluster
metallicity may be uncertain at a level of $\sim0.2$~dex, as indicated
by a series of papers which reached somewhat contrasing results for
these relations (see e.g. Pagel \& Tautvaisiene 1998; Dirsh et
al. 2000). Keeping the IMF fixed, the integrated $M_V$ changes with
metallicity \feh\ at a rate $\Delta M_V/\Delta\feh$ which is
approximately $\sim0.25$~mag/dex for ages lower than 1.2~Gyr, and
$\sim0.7$ otherwise\footnote{The change in $\Delta M_V/\Delta\feh$
at 1.2~Gyr is determined by the presence of the RGB and red clump at
later ages, whose integrated $V$-band light is more sensitive to \feh\
than the one from the main sequence.}. Therefore, errors of
$\sim0.2$~dex in the AMR would translate in errors of $\sim15~\%$ in
the derived lifetimes. Again, this is still smaller than the typical
errors caused by the poor statistics; on the other hand, it is also
evident that our results would benefit from a better accessment of the
AMRs in both the Magellanic Clouds.}

\section{Discussion and conclusions}
\label{sec_conclu}

Despite the large error bars, Table~\ref{tab_binned} and
Fig.~\ref{fig_lifetimes} clearly indicate that C-star lifetimes have
values of about $2$ to $3$ Myr, for stars in the mass interval from
$\sim1.5$ to $\sim2.8\,M_\odot$ (Fig.~\ref{fig_lifetimes}), and for
metallicities comprised between the $-0.3$ and $-0.7$~dex implied by
LMC and SMC data. There is also an indication that the peak of C-star
lifetime shifts to lower masses (from slightly above to slightly below
2~\Msun) as we move from LMC to SMC metallicities. The M-giant
lifetimes also peak at about 2~\Msun\ in the LMC, with a maximum value
of about 4 Myr. In the SMC the M-giant lifetimes appear much shorter,
but actually they are poorly constrained by present data.

These lifetimes correspond to the optically-visible TP-AGB phase. In
their study of IR sources, van Loon et al. (2005) find that
1.3--3~\Msun\ stars spend of the order of 10--20 percent of their AGB
lifetimes as optically-obscured, bright IR objects, with mass losses
higher than about $10^{-6}$~\Msun\,yr${-1}$. Assuming the typical
lifetimes above the RGB tip to be 1~Myr, they derive
1--2$\times10^5$~yr for the duration of this superwind
phase. Moreover, since about $3\times10^5$~yr is the time required for
loosing the envelope masses of the sample stars with the observed
(superwind) mass loss rates, van Loon et al. estimate that 30--70
percent of the mass loss of AGB stars occurs as superwind.  If we
repeat the same reasoning using our own estimates for the lifetimes in
the TP-AGB phase, which are at least twice larger than the value used
by van Loon et al. (2005), we find that the superwind phase may
account for {\em all} the mass loss during the TP-AGB phase. Of
course, this indication is anyway very uncertain because derived from
small numbers of stars.

\begin{figure}  
\resizebox{\hsize}{!}{\includegraphics{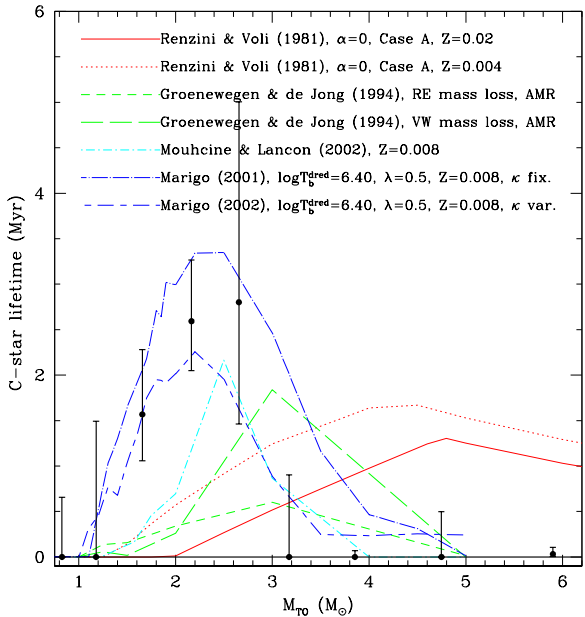}}
\caption{
The lifetimes of the C-star phase in the LMC as inferred from the
cluster data, as a function of turn-off mass (dots with error bars),
compared to the predictions of some TP-AGB models in the
literature. The comparison is made with models computed for the LMC
present metallicity ($Z\simeq0.008$), whenever available.}
\label{fig_models}
\end{figure}

We have also verified that several models in the literature present
C-star lifetimes significantly shorter than the values we find for the
LMC, as illustrated in Fig.~\ref{fig_models}. The comparison is shown
only for a subset of TP-AGB models which has already been used, with
various procedures, into population synthesis of galaxies.

These models can be roughly classified into two classes: the ones
previous to Groenewegen \& de Jong (1993) are {\em uncalibrated}, in
the sense that they fail in reproducing the C-star luminosities
observed in the Magellanic Clouds (as discussed for the first time by
Iben 1981). Renzini \& Voli (1981) TP-AGB models belong to this class;
as it can be seen in Fig.~\ref{fig_models}, they present too low
C-star lifetimes as compared to the data, and are systematically
shifted towards a much higher range of masses. Notice that we plot
just a subset of Renzini \& Voli's $\alpha=0$ models (their tables 1a
and 1f); their models with $\alpha>0$, which take into account the
effect of hot-bottom burning in more massive stellar envelopes, would
present even lower C-star lifetimes for $\mto\ga4$~\Msun. It is clear
that these models are highly discrepant with the data.

Groenewegen \& de Jong (1993) presented the first {\em calibrated}
TP-AGB models, in which the poorly known parameters determining the
occurrence and efficiency of third dredge-up events were tuned so as
to reproduce the LMC and SMC CSLFs. Indeed, all post-1993 models we
plot in Fig.~\ref{fig_models} have positive C-star lifetimes in the
right mass range, with the minimum mass for the present of C-stars
being close to 1.5~\Msun, and a peak lifetime located somewhere
between 2 and 3~\Msun, in rough agreement with the cluster data. This
agreement is just expected, since fitting the main features of the
CSLF requires imposing efficient dredge-up for masses as low as
1.4~\Msun, as discussed in detail by Groenewegen \& de Jong (1993) and
Marigo et al. (1999).

However significant discrepancies appear in the C-star lifetimes of
calibrated models too. For instance, lifetimes in Groenewegen
\& de Jong (1994) tend to be too short, with maximum values of just
0.6~Myr for models assuming Reimers' (1975) mass loss formula (RE in
the plot), and 1.8 Myr for Vassiliadis \& Wood's (1993) one (VW in the
plot). The discrepancies are more evident in the mass range between
1.5 and 2.5~\Msun, where model lifetimes fall to less than 0.5~Myr,
whereas the {ones derived here} are closer to 2~Myr. Notice that the
comparison of Fig.~\ref{fig_models} is not strictly correct because
Groenewegen \& de Jong (1994) tracks do not include overshooting,
whereas the LMC cluster data have been age-dated using overshooting
models. We estimate that, in order to be put Groenewegen \& de Jong
models in the same scale as the data, their masses should be reduced
by a factor of about 20 per cent. This would not solve the differences
in lifetimes.

The same discrepancy is shared, to a lower extent, by Mouhcine \&
Lan\c con's (2002) $Z=0.008$ models, which present the right peak
lifetimes of $\sim2.5$~Myr, but again have too low lifetimes at masses
$\la2$~\Msun. {Due to the significantly lower metallicities of the
old LMC clusters, and to the high metallicity dependence expected for
the C-type lifetime, this discrepancy for $M_{\rm TO}\la2$~\Msun\
would likely be reduced if the comparison was performed with models of
smaller metallicity (say $Z=0.004$).}  Notice that Mouhcine \& Lan\c
con (2002) adopt the same dredge-up parameters as Groenewegen \& de
Jong (1993), although {many} of their model ingredients have
changed. Therefore, the Mouhcine \& Lan\c con (2002) models are not
strictly calibrated on the CSLF, although their C-star luminosities
should not be much far from the observed ones.

Marigo (2001) models do not present these discrepancies in their
C-star lifetimes at lower masses, and have a peak lifetime of 3.4~Myr,
which is well compatible with the data. This could be considered as a
set of TP-AGB models which comply with both CSLF and lifetime
constraints. However, we know already that these models -- as well as
{\em all} of the previously mentioned ones -- are wrong for a
different reason: they do not consider the crucial effect of variable
molecular opacities as the chemical composition changes along the
TP-AGB evolution (see Marigo 2002). The effect of the variable
opacities is illustrated by the models labelled ``Marigo (2002)'' in
Fig.~\ref{fig_lifetimes}; it can be noticed that assuming the same
dredge-up parameters as calibrated by Marigo (2001), the change from
fixed (solar-scaled) to variable opacities ($\kappa$-fix and
$\kappa$-var cases, respectively) causes a reduction of
$\sim35$~percent in the C-star lifetimes. These would be still
compatible with the cluster data. However, the new models have also
lower mean luminosities and hence they fail to reproduce the CSLF. In
fact, updating of the input physics of TP-AGB models is not enough,
and a re-calibration of dredge-up parameters become necessary in this
case.

In conclusion, we have shown that present data for AGB stars in LMC
clusters represent useful -- and so far neglected -- constraints to
the lifetimes of TP-AGB models. Checking for these constraints should
be especially important if the TP-AGB models are to be used into
evolutionary population synthesis. In fact, using models with the
right luminosities (i.e. calibrated with the CSLF) but with too low
lifetimes would lead to underestimate the contribution of TP-AGB stars
to the integrated light of single-burst stellar populations.

In a following paper, we will present updated TP-AGB models computed
for the variable-opacity case, in which the lifetimes here derived are
adopted, together with the observed CSLF in the Magellanic Clouds, in
the calibration procedure of the main model parameters.



\end{document}